\documentclass[twocolumn,showpacs,preprintnumbers,amsmath,amssymb,prb]{revtex4}

\usepackage{epsfig}
\usepackage{graphicx}
\usepackage{dcolumn}
\usepackage{bm}
\usepackage{color}

\begin{document}

\title{Charge and spin Drude weight of the one-dimensional 
extended Hubbard model at quarter-filling}

\author{Tomonori Shirakawa}
\affiliation{
Graduate School of Science and Technology, 
Chiba University, Chiba 263-8522, Japan}
\affiliation{
Institut f\"ur Theoretische Physik, 
Leibniz Universit\"at Hannover, Appelstrasse 2, D-10607 Hannover, Germany}

\author{Eric Jeckelmann}
\affiliation{Institut f\"ur Theoretische Physik, 
Leibniz Universit\"at Hannover, Appelstrasse 2, D-10607 Hannover, Germany}

\date{\today}

\begin{abstract}
We calculate the charge and spin Drude weight of the one-dimensional 
extended Hubbard model with on-site repulsion $U$ and
nearest-neighbor repulsion $V$ at quarter filling 
using the density-matrix renormalization group method combined with a variational principle.
Our numerical results for the Hubbard model ($V=0$) agree with exact results obtained 
from the Bethe ansatz solution. 
We obtain the contour map for both Drude weights in the $UV$-parameter space for repulsive 
interactions. 
We find that the charge Drude weight is discontinuous across the 
Kosterlitz-Thouless transition between the Luttinger liquid and
the charge-density-wave insulator, while
the spin Drude weight varies smoothly and remains finite in both phases. 
Our results can be generally understood using bosonization and 
renormalization group results. 
The finite-size scaling of the charge Drude weight is well fitted by a 
polynomial function of the inverse system size in the metallic region.
In the insulating region we find an exponential decay of the finite-size 
corrections with the system size and a universal relation
between the charge gap $\Delta_c$ and
the correlation length $\xi$ which controls this exponential decay.
\end{abstract}


\maketitle

\section{Introduction}
The transport properties of low-dimensional strongly correlated electron 
systems are currently a subject of great interest because of recent 
experimental observations in quasi-one-dimensional materials
and because of their connection to the rapidly evolving field of 
nonequilibrium physics in strongly correlated quantum systems.\cite{zotos}
Most studies of transport properties in one-dimensional quantum many-body 
systems have been within the linear response theory.\cite{kubo} 
One fundamental quantity is the Drude weight defined as the 
zero-frequency contribution to the real part of the conductivity. 
Thus a finite Drude weight implies ballistic transport, and it has 
been proposed as a criterion for distinguishing metallic and  
insulating phases in a Mott transition.\cite{kohn}

Experiments on quasi-one-dimensional organic conductors\cite{jerome} 
show large deviations from the predictions of band theory. 
According to previous studies, the inter-site Coulomb repulsion plays a crucial 
role in these materials.\cite{seohotta} 
Therefore, the most simple effective model for their electronic properties is
a one-dimensional extended Hubbard model. 
In such one-dimensional models of interacting fermions, 
the quasiparticle concept breaks down, and 
the properties of the system do not resemble those of a Fermi liquid. 
Instead, low-energy excitations are made of 
independent elementary excitations for spins (spinons)
and charge (holons).\cite{lorenz, schulz}
Moreover, the space- and time-dependent correlation functions 
display unusual power-law decays. Their exponents are not universal but 
depend on the strength of the interaction. 
One-dimensional metallic systems belong to the generic class of 
Tomonaga-Luttinger liquids (TLL).\cite{emery, solyom, voit, giamarchibook} 
Their characteristic quantities are the so-called TLL parameters $v_{\rho}$, 
$v_{\sigma}$, $K_{\rho}$, and $K_{\sigma}$. 
Here $v_{\rho}$ and $v_{\sigma}$ are 
the velocity of charge and spin excitations, respectively, 
and  $K_{\rho}$ and $K_{\sigma}$ determine the 
algebraic decay of correlation functions. 

Recently, the discovery of the colossal magnetic heat transport 
in spin ladder materials such as (Sr,Ca,La)$_{14}$Cu$_{24}$O$_{41}$, 
where the magnetic contribution to the total thermal conductivity 
exceeds the phonon contribution substantially, 
has sparked interest in transport properties of quasi-1D spin models.
\cite{sologubenko, hess, kudo}
Understanding the transport properties of theoretical models 
is of great importance for the 
interpretation of transport or NMR measurements but it is still an open problem 
for quantum systems involving many coupled degrees of freedom.\cite{zotos, meisner}
Therefore, the development of methods for computing transport properties such as the Drude
weight in strongly correlated systems and the investigation of theses properties
are much anticipated. 

In this paper we study the charge and spin Drude weight of the 
one-dimensional extended Hubbard model at quarter filling 
using the density-matrix renormalization group (DMRG) method
with periodic boundary conditions. 
This model is known to be "non-integrable" for general values 
of the parameters\cite{poilblanc} and thus not amenable to an exact calculation
of the Drude weight contrary to the Hubbard model.\cite{stafford} 
Investigations based on the g-ology,\cite{solyom} 
bosonization,\cite{bosoni, yoshioka1, yoshioka2} 
and the renormalization group\cite{fourcade, andergassen} have provided analytic insight, 
particularly in the weak coupling regime. 
Both exact diagonalization\cite{edcal,milazotos,penc,nakamura,sano} calculations and 
quantum Monte Carlo simulations\cite{qmccal} have 
clarified a number of questions at intermediate and strong coupling.   
In the last decade the DMRG method has been
successfully used to investigate many properties of one-dimensional strongly correlated
lattice models\cite{schmitteckert, ejima}  
but a precise calculation of charge and spin Drude weights has not been reported yet. 
Therefore, we still lack a comprehensive picture of ballistic transport
in the one-dimensional extended Hubbard model. 

Our paper is organized as follows.  
In Sec.\ref{sec:model}, we summarize some properties of the 
extended Hubbard model and its ground state phase diagram and 
introduce the charge and spin Drude weight. 
In Sec.\ref{sec:method}, we explain the DMRG based method for 
calculating the Drude weight. 
Our results for the thermodynamic limit 
are presented in Sec.\ref{subsec:results} and 
the finite-size scaling is discussed in detail
in Sec.\ref{subsec:scaling}.
Finally, we  summarize our work in the last section.

\section{Model and Drude weight}
\label{sec:model}

We study the one-dimensional extended Hubbard model which is defined by the Hamiltonian
\begin{eqnarray}
H &=& H_t + H_U \\
H_t &=&  - t \sum_{l, \sigma} 
\left( c_{l,\sigma}^{\dagger}c^{\phantom{\dagger}}_{l+1,\sigma} + 
c_{l+1,\sigma}^{\dagger}c^{\phantom{\dagger}}_{l,\sigma}  \right) \\
H_U &=&  U \sum_{l} n_{l,\uparrow}n_{l,\downarrow} + V \sum_{l} n_{l} n_{l+1}
\label{eq:ham}
\end{eqnarray}
where $c_{l,\sigma}^{\dagger}$ ($c_{l,\sigma}$) is 
the creation (annihilation) operator for an electron with spin 
$\sigma$ ($=\uparrow,\downarrow$) at site $l=1,\dots,L$, 
$n_{l,\sigma}=c_{l,\sigma}^{\dagger}c_{l,\sigma}$ 
is the density operator, and $n_{l} = n_{l, \uparrow} + n_{l, \downarrow}$.
We use periodic boundary conditions throughout. 
$t > 0$ is the nearest-neighbor hopping integral along the chain, 
$U \geq 0$ is the onsite Coulomb interaction, and $V \geq 0$ is the 
nearest-neighbor Coulomb interaction. 
A quarter-filled band corresponds to a system with $N=L/4$ electrons of each spin,
a Fermi wavevector $k_{\mathrm{F}}=\frac{\pi}{4}$, and a Fermi velocity
$v_{\mathrm{F}}=2t\sin(k_{\mathrm{F}})=\sqrt{2}t$ in the Fermi gas ($U=V=0$).

\begin{figure}[tbp]
\begin{center}
\includegraphics[width=0.8\hsize]{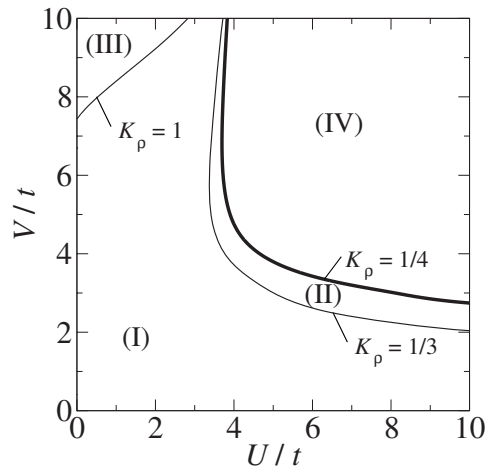}
\caption{The phase diagram of the $t$-$U$-$V$ model at quarter-filling 
determined from DMRG calculations in Ref.~\onlinecite{ejima}. 
In region (I), $1/3<K_{\rho}<1$ and
$2k_{\mathrm{F}}$-SDW correlations are dominant. 
Region (II) is characterized by dominant $4k_{\mathrm{F}}$-CDW correlation 
and $1/4<K_{\rho}<1/3$. 
In region (III), triplet pairing correlations dominate because 
of $K_{\rho}>1$. 
The region (I), (II), and (III) are metallic states while region 
(IV) is an insulating state with $4k_{\mathrm{F}}$ charge ordering. 
}
\label{fig:phasediagram}
\end{center}
\end{figure}

This model has been studied extensively by a variety of techniques. 
It is known to be "non-integrable" for general values 
of the parameters \cite{poilblanc} on the basis of energy level statistics 
although exact results can be obtained in three limits ($V=0$, $U= + \infty$, $V= + \infty$). 
For $V=0$, the model becomes the regular Hubbard model. At quarter-filling, 
it is known to be metallic \cite{lieb} with dominant 
2$k_{\mathrm{F}}$-spin-density-wave (SDW) fluctuations,  
and its low-energy excitations are of the TLL type. 
For $U=+\infty$, the quarter-filled electron model is equivalent to a half-filled 
spinless fermion model which upon increasing $V$ from zero 
undergoes a phase transition from a TLL phase to a $4k_{\mathrm{F}}$-charge-density-wave (CDW) 
insulator at $V = 2t$.\cite{halden} 
For $V=+\infty$, onsite electron pairs cannot move, while the unpaired electrons have the same 
kinetic energy as spinless fermions interacting with an infinitely strong
nearest-neighbor repulsion.
This system has a Bethe ansatz solution \cite{yangyang,fowler,milazotos,penc}. 
It is a $4k_{\mathrm{F}}$-CDW insulator for $U > U_c = 4t$
whereas it is phase separated for $U < U_c$.
In the weak-coupling limit ($U,V \ll t$), the model can be mapped onto a $g$-ology model
and investigated using bosonization and renormalization group techniques.
\cite{solyom, bosoni, yoshioka1, yoshioka2, fourcade, andergassen}

The ground-state phase diagram of the quarter-filled  extended Hubbard model
for repulsive interactions 
was first determined using exact diagonalizations.\cite{milazotos}
Recently, the precise ground-state phase diagram and the TLL exponent $K_{\rho}$ 
have been obtained for a  wide region of the $UV$-parameter space using the 
DMRG method.\cite{ejima}
These results are summarized in Fig.~\ref{fig:phasediagram}, where 
four different phases are represented: 
(I) A metallic phase with $1/3 \le K_{\rho} \le 1$ 
where the system has dominant 2$k_{\mathrm{F}}$-SDW fluctuations, 
(II) a metallic phase with $1/4 \le K_{\rho} \le 1/3$ 
where the system has dominant 2$k_{\mathrm{F}}$-CDW fluctuations, 
(III) a metallic phase ($K_{\rho} \ge 1$) 
where the system has dominant superconducting fluctuations, 
and (IV) an insulating phase (i.e. with a finite charge gap) where the system
has a long-range ordered $4k_{\mathrm{F}}$-CDW.
All four phases have gapless spin excitations.
\cite{milazotos,penc,nakamura} 
Finite spin gaps have been reported in previous exact diagonalization 
studies for large $V$ in phase (III) but our DMRG calculations indicate
that the spin gap vanishes in the thermodynamic limit at least for all 
$V \leq 10t$.
On the metal-CDW transition line $K_{\rho} = 1/4$.
It has been reported that this transition is of the Kosterlitz-Thouless 
type,\cite{nakamura} 
and higher order scattering processes, including the 8$k_{\mathrm{F}}$-Umklapp scattering 
via the upper band around $\pm 3k_{\mathrm{F}}$, 
play a crucial role in the weak-coupling theory of the phase diagram. 
\cite{schulz, suzumura, yoshioka1, yoshioka2}

Let us consider external fields $\phi_{\rho}$ and $\phi_{\sigma}$  which modify the kinetic energy 
\begin{eqnarray}
H_t (\phi_{\rho}, \phi_{\sigma} ) = 
-t \sum_{l,\sigma} \left( e^{i(\phi_{\rho}+\sigma\phi_{\sigma})/L} c_{l+1,\sigma}^{\dagger} c^{\phantom{\dagger}}_{l,\sigma} + \mathrm{h.c.} \right) .
\label{eq:external}
\end{eqnarray}
$\phi_{\rho}$ is the magnetic flux threading the system and $\phi_{\sigma}$ 
is the magnetic flux given by a fictitious spin-dependent vector potential.\cite{chandra} 
Time-dependent fields $\phi_{\mu}$ generate a charge ($\mu = \rho$) or spin  ($\mu = \sigma$) 
current $\langle J_{\mu}\rangle$ in the system.
Within the linear response theory  the charge and spin conductivities have the form
\cite{meisner, shastry, millis, fye, kopietz} 
\begin{eqnarray}
\mathrm{Re} \sigma_{\mu} (\omega) = \pi D_{\mu} \delta (\omega) 
+ \sigma_{\mu}^{\mathrm{reg}} (\omega) 
\end{eqnarray}
where $\sigma_{\mu}^{\mathrm{reg}}(\omega)$ is assumed to be 
regular at $\omega = 0$. 
The coefficient $D_{\rho}$ ($D_{\sigma}$) of the $\delta$-function $\delta (\omega)$ is 
called charge (spin) Drude weight for the ballistic transport, and is given by 
\begin{eqnarray}
D_{\mu} & = & S_{\mathrm{K}} + S_{\mu} \nonumber \\
& = & -\frac{1}{2L} \left< \psi_0 \right| H_t  \left| \psi_0 \right> - 
\frac{1}{L} \sum_n \frac{\left|\left< \psi_n \right| J_{\mu} 
\left| \psi_0 \right>\right|^2}{E_n - E_0}  ,
\label{eq:drude}
\end{eqnarray}
where $\psi_n$ denotes an eigenstate of $H$ with energy $E_n$ 
and the ground state corresponds to $n=0$. 
The first term (up to the prefactor $-1/2L$ ) is the total kinetic energy
while the second term (up to a prefactor $-\pi/2$) is the total spectral weight
of the incoherent part $\sigma_{\mu}^{\mathrm{reg}}(\omega)$ of the conductivity
and describes the reduction of the Drude weight caused by incoherent scattering processes.
The charge and spin current operators are $J_{\mu} = \sum_i j_{\mu,i} $
with local current operators defined from the continuity equation 
\begin{eqnarray}
i \left[ H , d_{\mu,l} \right] + j_{\mu,l+1} - j_{\mu,l} = 0, 
\end{eqnarray}
where $d_{\rho,l} = n_{l}$  and $d_{\sigma,l} =  n_{l,\uparrow}-n_{l,\downarrow}$. 
In our model, the precise form of $j_{\mu,l}$ is
\begin{eqnarray}
&{}& j_{\rho,l} = -i t \sum_{\sigma} c_{l+1,\sigma}^{\dagger}c^{\phantom{\dagger}}_{l,\sigma} + \mathrm{h.c.} \\
&{}& j_{\sigma,l} = -i t \sum_{\sigma} \sigma \left ( c_{l+1,\sigma}^{\dagger}c^{\phantom{\dagger}}_{l,\sigma} + \mathrm{h.c.} \right )  .
\end{eqnarray}

The above spin current operator is different from the one used in spin models.
\cite{meisner, kopietzheisen}
Nevertheless, the spin Drude weights defined for an electron system or for a spin system 
have the same physical meaning because they characterize the response of the spin degrees of
freedom to the same external perturbation. 
The spin Drude weight $D_{\sigma}$ is thus defined as 
a precise analog of the charge Drude weight $D_{\rho}$. 
Therefore, a value $D_{\sigma}>0$ simply means that the system 
is an ideal spin conductor, so that the spin transport is not diffusive.\cite{kopietz} 

Obviously, charge and spin Drude weight are equivalent in the non-interacting electron gas
($U=V=0$) because the second term of (\ref{eq:drude}) vanishes. 
Assuming that the low-energy excitation of our model can be expressed by the TLL theory, 
the charge and spin Drude weight can be represented by\cite{giamarchibook}
\begin{eqnarray}
D_{\mu} = u_{\mu} K_{\mu}/\pi
\end{eqnarray}
where $v_{\rho}$ ($v_{\sigma}$) is the renormalized 
charge (spin) velocity and 
$K_{\rho}$ ($K_{\sigma}$) is the renormalized 
TLL exponent of charge (spin) mode. 
In our model, the renormalized value of $K_{\sigma}$ is $K_{\sigma} = 1$ 
because the system has a SU(2) spin-symmetry and there is no spin gap
in the repulsive parameter region. Thus, $\pi D_{\sigma}=v_{\sigma}$. 
On the other hand, the behavior of $D_{\rho}$ is more complicated 
because $v_{\rho}$ and $K_{\rho}$ depends on the interaction parameters. 

\section{DMRG methods for Drude weights}
\label{sec:method}
A method for calculating the Drude weight (\ref{eq:drude}) with DMRG
was introduced several years ago \cite{schmitteckert}  but has been rarely used until now.
In this section we briefly summarize our implementation of this numerical method 
and discuss some technical details. 

The first term of eq.(\ref{eq:drude}) 
can be easily calculated using the ground state DMRG method. 
The second-term can be calculated by targeting the correction vector
$| \psi \rangle$ which is solution of
\begin{eqnarray}
\left( H - E_0 \right) | \psi \rangle = J_{\mu} | \psi_0 \rangle.
\label{eq:corvec}
\end{eqnarray}
The best implementation of this idea is a variational principle similar to the one
used for the calculation of dynamical correlation functions.\cite{eric2002,book08}
One considers the functional 
\begin{eqnarray}
W_{\mu} ( \psi ) = 
\langle \psi | \left( H - E_0 \right) | \psi \rangle 
- \langle J_{\mu} |  \psi \rangle 
- \langle \psi  | J_{\mu} \rangle. 
\label{eq:var1}
\end{eqnarray}
If the ground state is not degenerate, this functional has a unique minimum for 
the quantum state which is the solution of~(\ref{eq:corvec}).
It is easy to show that the value of the minimum is related to the 
second term of eq. (\ref{eq:drude}): 
\begin{eqnarray}
W_{\mu}(\psi_{\mathrm{min}}) = - \sum_n 
\frac{ \left| \left< \psi_n \right| J_{\mu} \left| \psi_0 \right> \right|^2}{E_n-E_0}. 
\label{eq:var2}
\end{eqnarray}
Our method consists in calculating the ground state and then minimizing this functional
with DMRG.
Note that this approach does not work
for systems with a degenerated ground state. 
Therefore, we always choose appropriate system sizes $L$ and numbers of electrons $N$
to get a nondegenerate ground state. 

Another approach for obtaining the Drude weight  with DMRG is to compute  
the dynamical current-current correlation function 
\begin{eqnarray}
C_{\mu,\eta} \left( \omega \right) = - \left< \psi_0 \right| 
J_{\mu} \frac{1}{\omega + E_0 - H + i \eta} J_{\mu} \left| \psi_0 \right>
\label{eq:ddmrg}
\end{eqnarray}
using the dynamical DMRG (DDMRG) method.\cite{eric2002,book08}
The imaginary-part of this quantity satisfies 
\begin{eqnarray}
\frac{1}{\omega} \lim_{\eta \to 0} \mathrm{Im} C_{\mu,\eta} \left( \omega \right) 
= \sigma_{\mu}^{\mathrm{reg}} (\omega)  ,
\label{eq:ddmrg2}
\end{eqnarray}
which has been previously used to study the optical absorption of various one-dimensional 
insulators including the extended Hubbard model at half filling.\cite{eric2003}
The real part of the correlation function yields 
\begin{eqnarray}
\lim_{\eta \to 0} \mathrm{Re}C_{\mu,\eta}(0) = \sum_n 
\frac{\left|\left<\psi_n \right| J_{\mu} \left| \psi_0 \right>\right|^2}{E_n-E_0} ,
\end{eqnarray}
which can be expected from the Kramers-Kronig relation 
with the f-sum rule of conductivity. Therefore,
one can in principle calculate the second-term of (\ref{eq:drude}) 
using DDMRG. 
However, a DMRG calculation using (\ref{eq:var1}) and (\ref{eq:var2}) is faster and more accurate
than a DDMRG calculation of (\ref{eq:ddmrg}) and (\ref{eq:ddmrg2}). 
In the first approach the error in the value of 
the minimum $W_{\mu} \left( \psi_{\mathrm{min}} \right)$ is of the order of 
$\epsilon^2$  if we can calculate target states with an error of the order 
$\epsilon \ll 1$ within DMRG.  With the DDMRG method the error in the real part of 
$C_{\mu,\eta}(\omega)$ is of the order of $\epsilon$
(see the discussion in~Refs.~\onlinecite{eric2002,book08}).

As originally noted by Kohn,~\cite{kohn} the Drude weight (\ref{eq:drude}) 
can be calculated from 
the dependence of the ground state energy on the applied field $\phi_{\mu}$ using 
\begin{eqnarray}
D_{\mu} = L \left. \frac{\partial^2 E_0 (\phi_{\mu})}{\partial \phi^2_{\mu}}
\right|_{\phi_{\mu}=0}  .
\label{eq:difdrude}
\end{eqnarray}
Therefore, one can calculate the Drude weight with DMRG using Eq.(\ref{eq:difdrude}). 
This approach has been demonstrated on the spinless fermion 
model\cite{schmitteckert,carvalho} but it requires treating complex Hamiltonians
and performing a numerically delicate second derivative of the ground-state energy 
with respect to $\phi_{\mu}$. 
Therefore, we have chosen the approach based on the 
variational principle~(\ref{eq:var1}) and the equation~(\ref{eq:var2}). 

\section{Results}
\label{sec:results}

We have carried out DMRG calculations for quarter-filled
chains with periodic boundary conditions and lengths up to $L=60$. 
The investigated system lengths are given by $L=8l+4$ with 
integers $l>1$ so that the number of electrons of each spin
($N = L/4$) is odd. Thus the ground state has momentum $P = 0$ and is
not degenerate.
We have kept up to $m \approx 3200$ density-matrix eigenstates 
in the DMRG procedure. The discarded weights are typically of 
the order $10^{-6} \sim 10^{-8}$ and the ground-state energy accuracy is
$\sim 10^{-4}t$.  All energies and physical quantities 
are extrapolated to the limit $m \to \infty$.  

\subsection{Thermodynamic limit}
\label{subsec:results}
\begin{figure}[tbp]
\begin{center}
\includegraphics[width=0.8\hsize]{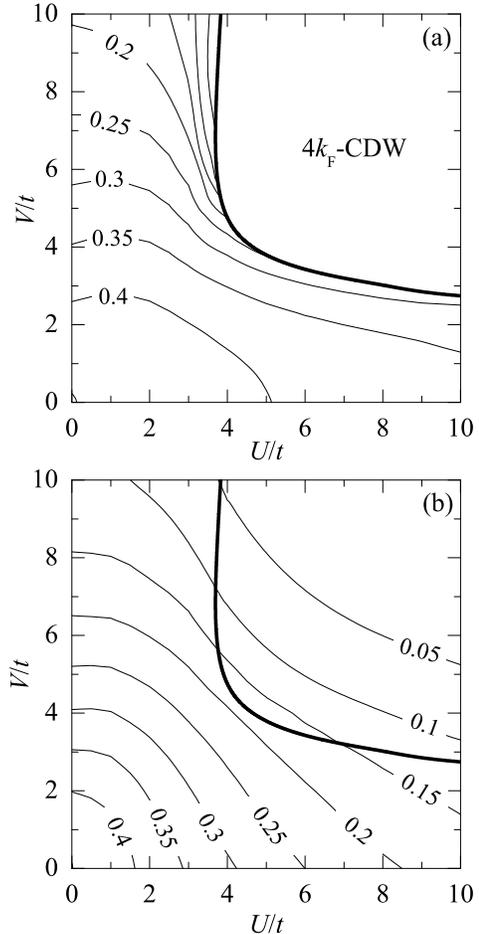}
\caption{Contour map for (a) the charge Drude weight $D_{\rho}$ and 
(b) the spin Drude weight $D_{\sigma}$ 
in the $UV$-parameter space of the extended Hubbard model at quarter filling.
The bold line represents the boundary of the metal-insulator transition 
determined from $K_{\rho}$ in Ref.~\onlinecite{ejima}. }
\label{fig:phase}
\end{center}
\end{figure}

We first discuss the DMRG results extrapolated to the thermodynamic limit ($L \to \infty$). 
The finite-size scaling is discussed in the next subsection.
Contour maps of the charge and spin Drude weights
are shown in Fig.~\ref{fig:phase}. 
We can summarize our main results in five points.
(i) Both $D_{\rho}$ and $D_{\sigma}$ have 
their maximum $D_{\mu}=v_{\mathrm{F}}/\pi = \sqrt{2}t/\pi$ 
at the noninteracting point 
($U=V=0$) and decrease monotonically as a function of increasing
$U$ and $V$.  (ii) Excepted for the noninteracting point, 
we observe a difference between $D_{\rho}$ and $D_{\sigma}$. 
(This is due to the well-known spin-charge separation, 
which is a typical property in one-dimensional systems.)
(iii) The charge Drude weight has no linear correction around the noninteracting point 
$\partial D_{\rho}/\partial U = \partial D_{\rho}/\partial V = 0$ at $U=V=0$. 
(iv) The charge Drude weight is larger than the spin Drude weight in the metallic phase. 
(v) $D_{\rho}$ seems to be discontinuous at the metal-insulator transition. 
To understand these feature we will discuss the behavior of
both Drude weights along the lines $V=0$, $U=0$, and $U=10t$ in more detail below. 

\begin{figure}[tbp]
\begin{center}
\includegraphics[width=\hsize]{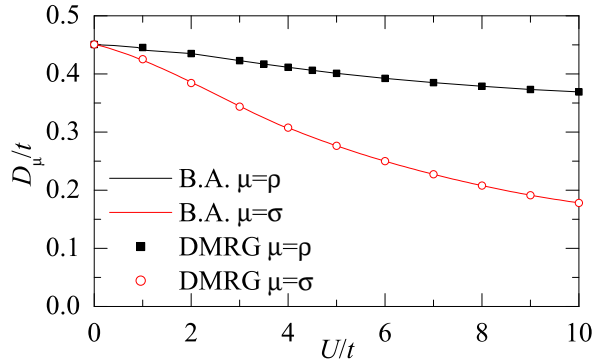}
\caption{Charge Drude weight $D_{\rho}$ and 
spin Drude weight $D_{\sigma}$ 
as a function of $U$ in the Hubbard model ($V=0$). 
Lines show the exact results from the Bethe ansatz solution. }
\label{fig:v00}
\end{center}
\end{figure}

Our DMRG calculation results for $V=0$ 
can be compared with the exact Drude weights $D_{\mu}^{\mathrm{exact}}$
obtained from the Bethe ansatz solution of the Hubbard model\cite{stafford}
as shown in Fig.~\ref{fig:v00}. 
Relative errors $\left(D_{\mu}^{\mathrm{exact}}-D_{\mu}^{\mathrm{DMRG}}\right)/D_{\mu}^{\mathrm{exact}}$ 
are below $10^{-4}$ for each system size $L \leq 60$ using up to 3000 density-matrix states. 
We note that the charge Drude weight is larger than the spin Drude weight
for all $U > 0$ in Fig.~\ref{fig:v00}.
The reduction of both Drude weights for finite interactions
can be understood qualitatively.  
The kinetic energy term $S_{\mathrm{K}}$ in eq.(\ref{eq:drude}) is
maximal for non-interacting electrons ($U=V=0$)
and decreases monotonically when $U$ (or $V$) increases. 
The second term $S_{\mu}$ in eq.(\ref{eq:drude}) equals 0 at the 
noninteracting point
and can only decrease to negative values for $U > 0$ (or $V > 0$).
(Note that $S_{\mu}$ is not a monotonic function of $U$ and $V$.
It has a minimum at finite interactions as it converges to zero 
in the strong-coupling limit.) 
Therefore, the decrease of the Drude weight is 
due to both the suppression of the kinetic energy
and the appearance of scattering processes.

In Fig.~\ref{fig:v00} we can see in both our DMRG results and the
exact Bethe ansatz results that $D_{\rho}$ has no linear correction in $U$ 
close to the noninteracting point ($U=0$),
whereas $D_{\sigma}$ seems to have a linear correction in $U$. 
This weak-coupling behavior has already been discussed.\cite{schulz, giamarchi1, giamarchi2}
The renormalized value of $K_{\sigma}$ is always $K_{\sigma}=1$
because there is no spin gap for the parameters investigated here.
Therefore, the lowest order correction to $D_{\sigma}$ yields
$D_{\sigma} = v_{\mathrm{F}}/{\pi} - U/(2 \pi^2 v_{\mathrm{F}})$, 
which is also consistent with the Bethe ansatz result.\cite{chubukov} 
The behaviour of $K_{\rho}$ is 
more complicated.\cite{schulz, ejima}
There is no first order correction from the interaction $U$ and thus
if we neglect the irrelevant incommensurate $4k_{\mathrm{F}}$-Umklapp 
scattering\cite{giamarchi1} 
and higher order corrections in $U$,\cite{yoshioka1}
$D_{\rho} = v_{\rho}K_{\rho}/\pi = v_{\mathrm{F}}/\pi$. 
To understand the decrease of $D_{\rho}$, 
the effect of the irrelevant $4k_{\mathrm{F}}$-Umklapp scattering has 
to be taken into account. 
This correction is second or higher order in the interaction
and explains the non-linear decrease of $D_{\rho}$. 
Note that, if we neglect the irrelevant $4k_{\mathrm{F}}$-Umklapp scattering but 
take higher order corrections in $U$ into account (see Ref.\onlinecite{yoshioka1}), 
$D_{\rho}$ increases, which contradicts both our numerical results and the Bethe ansatz results. 
Thus, though those higher order corrections are important 
for qualitatively understanding the phase diagram of this model, 
they are not sufficient for a quantitative analysis.

In the strong coupling limit $U\to \infty$ we expect the following behavior:
$D_{\rho}$ approaches the value $t/\pi$ because $v_{\rho} \to 2t\sin 2k_{\mathrm{F}} = 2t$
and $K_{\rho} \to 1/2$
while $D_{\sigma}$ goes to 0 because $v_{\sigma} \sim O(1/U)$.
Therefore, along the $V=0$ line, $D_{\rho}$ is larger than $D_{\sigma}$ in both 
the $U \to 0$ and $U\to \infty$ limits and our DMRG results show that this
relation holds also for all finite $U \geq 0$. 

\begin{figure}[tbp]
\begin{center}
\includegraphics[width=\hsize]{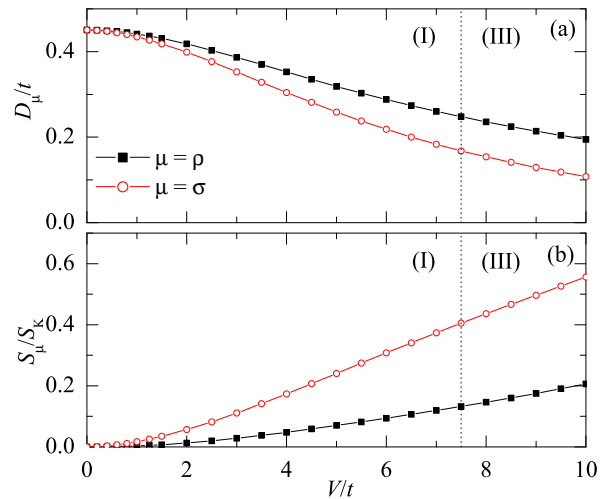}
\caption{(a) The charge Drude weight $D_{\rho}$ and 
spin Drude weight $D_{\sigma}$ 
as a function of $V$ along the $U=0$ line. (b) The ratio $S_{\mu}/S_{\mathrm{K}}$ where 
$S_{\mathrm{K}}$ and $S_{\mu}$ are the first term and second term in 
eq.(\ref{eq:drude}), respectively. 
Vertical lines show the boundary between the Luttinger liquid phases with
dominant SDW (I) and pairing (III) fluctuations.
Other lines are guides for the eyes.
}
\label{fig:u00}
\end{center}
\end{figure}

Results for the $U=0$ line of the $UV$-parameter space are 
shown in Fig.~\ref{fig:u00}(a).
Again we clearly see that both Drude weights decrease monotonically
with increasing interaction and that the spin Drude weight is less than 
the charge Drude weight for all $V > 0$. 
In that case, however, it seems that both Drude weights 
have no linear correction in $V$ at the noninteracting point $(U=V=0)$.
This is consistent with the weak-coupling theory which yields
no correction to $v_{\rho}K_{\rho}$ or $v_{\sigma}$
in first order in $V$.\cite{giamarchibook}
The faster reduction of $D_{\sigma}$ with increasing $V$ is due
to the much stronger incoherent scattering processes for spin excitations
than for charge excitations,
which is demonstrated by the larger second term $S_{\mu}$ of
eq.(\ref{eq:drude}) as shown in Fig.~\ref{fig:u00}(b). 

We also find that $S_{\mu}/S_{\mathrm{K}}$ increases in the Luttinger liquid 
phase (III) with dominant superconducting correlations, 
which occurs for $V > 7.5t$.
This is an unusual behavior for this ''superconducting'' phase 
because $S_{\rho}/S_{\mathrm{K}}=0$ for a true superconductor.\cite{scalapino} 
In the $V\rightarrow\infty$ limit, both Drude weights go to zero 
because the system becomes phase-separated at $V=\infty$ for $U=0$.
The lattice is decomposed in finite-size domains of singly occupied or empty 
sites with an average electronic density $< 1/2$. 
These domains are separated by impenetrable immobile walls 
(the doubly occupied sites). There is a finite density of such pairs
for small $U$
because they reduce the average density on the other sites below $1/2$
and thus allow them to gain kinetic energy.
The charge Drude weight is zero despite the finite kinetic energy
[i.e., $S_{\mathrm{K}}\neq 0$ in (\ref{eq:drude})]
because charge motion is confined to finite domains by the infinite walls.\cite{rigol}
Therefore, there are only incoherent contributions to the charge conductivity
and $S_{\rho} = - S_{\mathrm{K}}$.
The spin Drude weight must also vanish for this reason and also because
the infinite nearest-neighbor interaction $V$ prohibits the formation
of any nearest-neighbor electron pairs and thus the 
effective magnetic interaction is zero.\cite{milazotos,penc}

\begin{figure}[tbp]
\begin{center}
\includegraphics[width=\hsize]{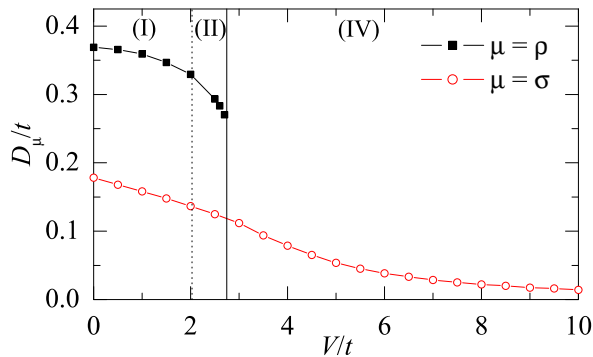}
\caption{The charge Drude weight $D_{\rho}$ and 
spin Drude weight $D_{\sigma}$ 
as a function of $V$ along the $U=10t$ line. 
The solid vertical line is the phase boundary between the metallic and
insulating (IV) states determined by the calculation of $K_{\rho}$ in 
Ref.~\onlinecite{ejima}. 
The dashed vertical line marks the boundary between the Luttinger liquid
regions with dominant SDW (I) and CDW (II) fluctuations.}
\label{fig:U10}
\end{center}
\end{figure}

Exact diagonalization studies\cite{milazotos} have shown that
the compressibility $\kappa_{\rho}$ increases with $V$ in 
this ''superconducting'' region and seems to diverge when
approaching the phase separation regime for $V \rightarrow \infty$.
As within the TLL approach the compressibility can be characterized by 
$\kappa_{\rho} = 2K_{\rho}/\pi v_{\rho}$, 
the divergence of $\kappa_{\rho}$ has been interpreted as a divergence
of $K_{\rho}$. 
However, our DMRG calculations show that 
$D_{\rho} = v_{\rho} K_{\rho}/\pi$ decreases as 
$V$ becomes very large. Thus $v_{\rho}$ goes to zero faster than
$K_{\rho}$ diverges toward $\infty$ (if it diverges) 
and we conclude that the divergence of $\kappa_{\rho}$  
is mostly due to the vanishing of the charge velocity
for $V \to \infty$.  
Unfortunately,
it is difficult to compute $K_{\rho}$ using 
the DMRG method close to the phase separation regime. Thus we
cannot determine whether $K_{\rho}$ diverges to $\infty$ or converges
to a finite value.  

\begin{figure}[tbp]
\begin{center}
\includegraphics[width=\hsize]{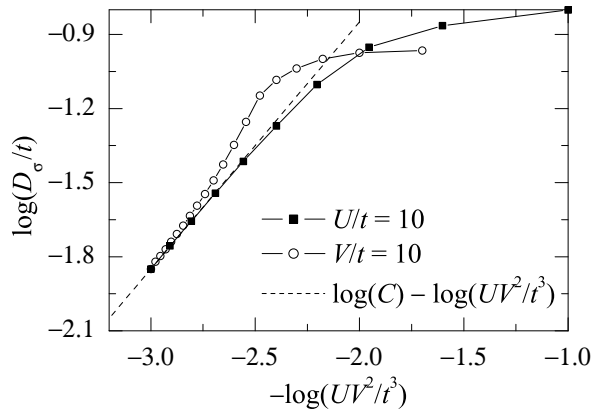}
\caption{Spin Drude weight $D_{\sigma}$ vs $-\log(UV^2/t^3)$ 
in the strong coupling region. Solid lines are guides for the eyes.}
\label{fig:dsvsjef}
\end{center}
\end{figure}

Our results for the $U = 10t$ line are shown in Fig.~\ref{fig:U10}.
As $V$ increases, there is a phase transition from the metallic state
to the $4k_{\mathrm{F}}$-CDW insulating state at 
$V_c/t \approx 2.74$.\cite{ejima}
In the metallic phase, both Drude weights decrease when $V$ increases
and we always find $D_{\rho} > D_{\sigma}$ as discussed previously.
The spin Drude weight changes smoothly and remains finite for all
values of the interaction parameters that we have investigated.
As seen in Fig.~\ref{fig:U10}
the charge Drude weight approaches a finite value when
$V \rightarrow V_c$ from the metallic side. As $D_{\rho}=0$
in the insulating phase for $V > V_c$, we conclude that
the charge Drude weight jumps from a finite value to zero
at the metal-insulator transition. Therefore, $D_{\rho}=0$ is
a discontinuous function of the interaction parameters.
[Note that the finite-system $D_{\rho}(L)$ calculated with DMRG are
smooth functions of the parameters and from the finite-size scaling
analysis alone one cannot determine whether $D_{\rho}$ has a jump
at $V=V_c$, see the next subsection.]
The transition from the metallic to the insulating phase is believed
to be of the Kosterlitz-Thouless type and to be caused
by the $8k_{\mathrm{F}}$-Umklapp scattering.\cite{yoshioka1,yoshioka2} 
This scattering is different 
from the usual $4k_{\mathrm{F}}$-Umklapp scattering 
which reduces the charge velocity $v_{\rho}$ in
renormalization group calculations and is always irrelevant.\cite{giamarchi1,giamarchi2}
Therefore, both $v_{\rho}$ and $K_{\rho}$ are renormalized to a finite value 
when approaching the phase boundary from the metallic side in the renormalization group analysis.

Unfortunately, our simulations show that it is difficult to determined the 
phase boundary using the discontinuity in $D_{\rho}$. We believe that 
the direct calculation of $K_{\rho}$ from correlation functions\cite{ejima}
is a more efficient approach for determining phase boundaries within DMRG
computations because the extrapolation to infinite system size is 
less difficult than for the Drude weights. 
This will be discussed in the following subsection.

In the strong-coupling limit of the insulating CDW phase (IV)
electrons are localized and there is an effective
Heisenberg-like interaction between their spins.
The effective exchange interaction $J_{\mathrm{eff}}$ 
between nearest-neighbor spins
can be derived from perturbation theory  and we obtain
up to fourth order $J_{\mathrm{eff}} \propto t^4/UV^2$. 
In a Heisenberg model $D_{\sigma}$ is proportional to 
the exchange interaction. 
Therefore we expect $D_{\sigma} = C t^4/UV^2$ in the strong-coupling limit
of our electronic model, where $C$ is an unknown  constant. 
In Fig.~\ref{fig:dsvsjef} we show our numerical results for the
strong-coupling limit on a double logarithmic scale.
One can clearly see a linear asymptotic behavior 
$\log(D_{\sigma}/t) = \log(C) -\log(UV^2/t^3)$ 
for large $U/t$ and $V/t$ in agreement with the strong-coupling analysis.

\subsection{Finite-size scaling}

With DMRG we have been able to investigate much larger system sizes than
in studies based on the exact diagonalization
method.~\cite{edcal,milazotos,penc,nakamura,sano}
Finite-size effects are thus smaller than in these previous studies
and in most cases we can perform
a reliable finite-size scaling analysis of DMRG results
for finite chains with periodic boundaries and extrapolated
to the limit of infinitely long chains.
Nevertheless, there are some difficulties close to phase boundaries as
usual and it is always informative to examine finite-size corrections.

\begin{figure}[tbp]
\begin{center}
\includegraphics[width=\hsize]{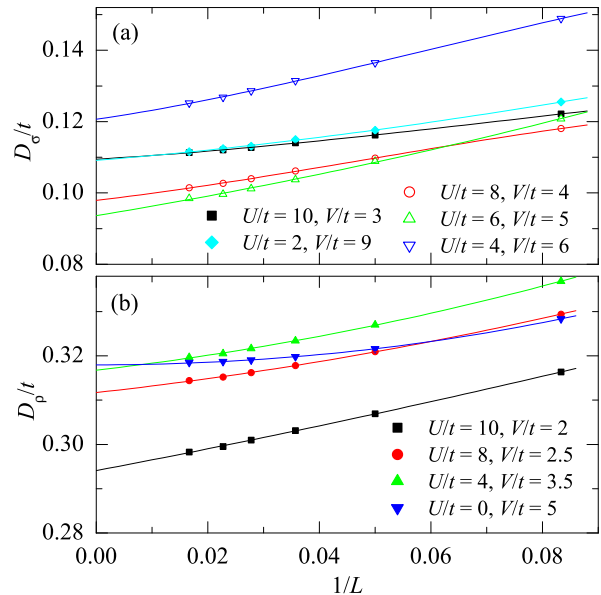}
\caption{(a) Spin Drude weight $D_{\sigma}$ 
as a function of inverse system length $1/L$ for several parameter sets
in the metallic phase (filled symbols) and in the insulating phase
(open symbols). Lines are polynomial fits. 
(b) Charge Drude weight $D_{\rho}$ for several parameter sets in the
metallic phase. Lines are polynomial fits.}
\label{fig:metal}
\end{center}
\end{figure}

Figure~\ref{fig:metal}(a) shows the spin Drude weight $D_{\sigma}(L)$ 
as a function of inverse system length $1/L$ for several parameter sets. 
We note that $D_{\sigma}(L)$ decreases monotonically and smoothly with increasing $L$
and also that it converges to a finite value and seems to be a convex function of $1/L$ 
for $L\rightarrow \infty$
both in the metallic and the insulating phase. 
This finite-size scaling reflects the absence of gap 
in the spin sector and the complete separation of low-energy 
spin and charge excitations in the quarter-filled extended Hubbard model for $U,V \geq 0$.
We can extrapolate the Drude weight to the thermodynamic limit systematically 
by performing a polynomial fit in $1/L$. We find that
the extrapolation is always very well behaved as seen from Fig.~\ref{fig:metal}(a). 
Especially, the relative errors are small, i.e. the difference between the 
extrapolated value $D_{\sigma}$ and the value of $D_{\sigma}(L)$ for the largest 
system size $L$ computed with DMRG is small compared to $D_{\sigma}$.

\label{subsec:scaling}

Figure.~\ref{fig:metal}(b) shows the charge Drude weight $D_{\rho}(L)$ 
as a function of inverse system length $1/L$ for several parameter sets 
in the metallic phase. As expected $D_{\rho}(L)$ converges to a finite
value for $L\rightarrow \infty$. As for the spin Drude weight
we find that finite-size corrections are positive and depend
smoothly on $L$ and that $D_{\rho}(L)$ seems to be a convex function
of $1/L$ for $L\rightarrow \infty$. 
Again such a finite-size scaling reflects the absence of gap in the charge 
sector in this part of the phase diagram.
We can extrapolate $D_{\rho}(L)$ to the thermodynamic limit 
systematically using a polynomial fit in $1/L$. This extrapolation is
very accurate as demonstrated by the comparison with the exact results
obtained from the Bethe Ansatz for the Hubbard model (see Fig.~3).
We note, however, that leading finite-size corrections are linear
in $1/L$ in the $U > V > 0$ region of the parameter space whereas
they are of the order of $1/L^2$ in the Hubbard model.~\cite{stafford}
In the $V > U$ region we find that finite-size corrections to $D_{\rho}$
are much smaller than for $U > V$ region and that the leading term
seems to be again of the order $L^{-2}$ as can be seen for $U=0,V=5t$
in Fig.~\ref{fig:metal}(b). 

\begin{figure}[tbp]
\begin{center}
\includegraphics[width=\hsize]{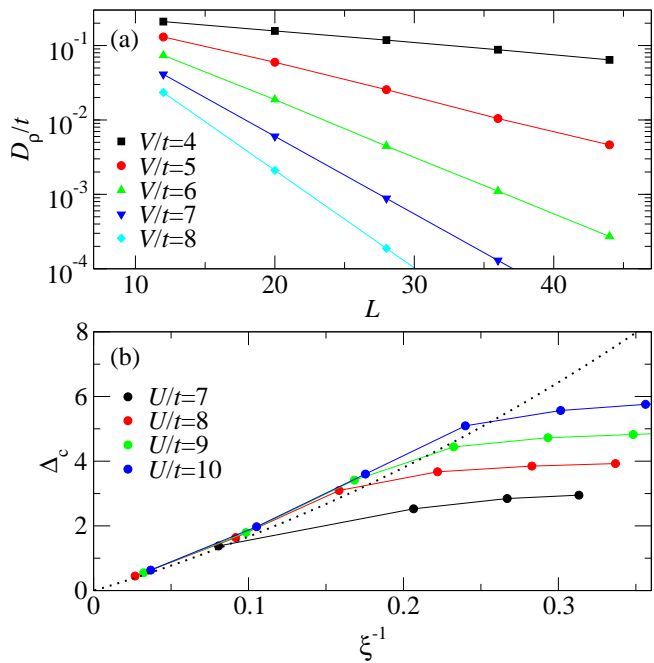}
\caption{(a) Semilog plot for the charge 
Drude weight $D_{\rho}$ 
at $U/t = 10$ as a function of the system size $L$. Lines are exponential fits.
(b) Charge gap $\Delta_c$ vs. inverse correlation length $\xi^{-1}$
for $U/t = 10, 9, 8, 7$ and varying $V$. Solid lines are guides to the eyes.
The dashed line is obtained from the Bethe ansatz solution~\cite{shastry}
in the limit $U/t=\infty$ as $V$ approaches $2t$ from above. 
}
\label{fig:inslator}
\end{center}
\end{figure}

In Fig.~\ref{fig:inslator}(a) we show the charge Drude weight $D_{\rho}(L)$ 
in the insulating phase
as a function of the system length $L$ for several parameter sets. 
The linear behaviour observed in this semilog plot indicates
that $D_{\rho}(L)$ decreases exponentially with increasing $L$
as in the Hubbard model at half filling.\cite{stafford}
The correlation length $\xi$ which characterizes this exponential decay,
$\log[D_{\rho}(L)/t] = -L/\xi$, depends upon $U$ and $V$. 
(We set the lattice constant $a=1$.) 
Near the metal-insulator phase boundary one expects 
that $\xi^{-1}$ should vanishes as the charge gap $\Delta_c$.
In particular, in the half-filled Hubbard model it has been shown that
$\xi^{-1}=\Delta_c/2t = \Delta_c/v_F$ in the weak coupling limit $U \ll t$.~\cite{stafford}
We have determined the correlation length $\xi$ in our model
from the slopes of $\log[D_{\rho}(L)/t]$ versus $L$.
The charge gap is obtained as the difference of 
ground-state energies extrapolated to the thermodynamic limit
\begin{eqnarray}
&&\Delta_c = \lim_{L \to \infty} \Delta_c(L), \\
&&\Delta_c(L) = \left[ E_L(1,1)+E_L(-1,-1) - 2 E_L(0,0) \right]/2 . \nonumber
\end{eqnarray}
Here $E_L(N_{\uparrow},N_{\downarrow})$ denotes the ground-state 
energy of a chain of length $L$ with $N_{\uparrow}$ up-spin and 
$N_{\downarrow}$ down-spin electrons added or removed from the quarter-filled band,
which can be easily computed using the DMRG method. 
Figure~\ref{fig:inslator}(b) shows $\Delta_c/t$ vs $\xi^{-1}$
for several values of $U$ and $V$. 
We see that $\xi^{-1}$ varies linearly with $\Delta_c$ for small gaps while 
deviations are apparent when the gap is large. 
The product $\xi \Delta_c$ tends to a universal value 
$\xi \Delta_c \approx 13.8 t \approx 9.76 v_F$
for $\Delta_c \rightarrow 0$ independently of $U$ and $V$.
This agrees with the relation between $\xi$ and $\Delta_c$ derived from Bethe ansatz 
results~\cite{shastry} for the effective spinless fermion model which corresponds to
our model in the limit $U/t=\infty$ as $V$ approaches $2t$ from above (see the dashed
line in fig.~\ref{fig:inslator}(b)).

\begin{figure}[tbp]
\begin{center}
\includegraphics[width=\hsize]{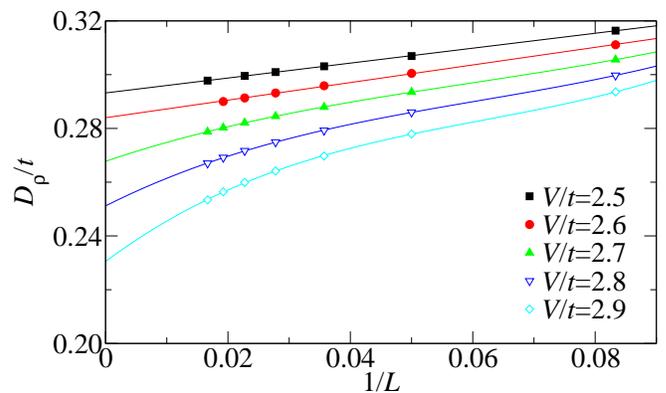}
\caption{(a) The charge Drude weight $D_{\rho}$ 
as a function of inverse system length 
$1/L$ at $U/t=10$ for various $V$ in the metallic phase (filled symbols, 
$V < V_c\approx 2.74t$)
and in the insulating phase (open symbols, $V > V_c$). 
Lines are polynomial fit. 
}
\label{fig:vicvc}
\end{center}
\end{figure}

It is also necessary to discuss the finite-size scaling of
the charge Drude weight in the vicinity of the metal-insulator
transition in more detail. 
In Fig.~\ref{fig:vicvc} we plot the charge Drude weight $D_{\rho}(L)$ as a function of 
inverse system length $1/L$ for $U=10t$ and several values of $V$ close to the
critical value $V_c \approx 2.74t$ determined in Ref.~\onlinecite{ejima}.
If we perform a polynomial fit in $1/L$, all results extrapolate to finite 
values for both phases while an exponential fit is meaningless.
Thus from our numerical results alone we would conclude that $D_{\rho}$
decreases smoothly as a function of $V$ through the critical value
$V_c\approx 2.74t$ determined in Ref.~\onlinecite{ejima}
and vanishes only for $V > V^{\ast}$, where $V^{\ast}$ is clearly
larger than $V_c$.

This failure of our finite-size analysis is easy to understand.
Obviously, 
if we assume a scaling behavior $D_{\rho}(L) \approx A \exp(- L/\xi)$ 
in the critical region $V \approx V_c$ of the insulating phase and $\xi$ diverges
as $V\rightarrow V_c$,
we need to treat exponentially large systems to observe the
correct finite-size scaling of $D_{\rho}(L)$.
Therefore, one has to use much larger sizes than the ones
used here (up to $L=60$) to determine the phase boundary from the extrapolated
charge Drude weights.
Unfortunately, it is very difficult to obtain accurate
results for larger periodic systems using DMRG.
We note that $D_{\rho}(L)$ seems to be a convex function of 
$1/L$ in the metallic phase but a concave function in the insulating
phase. Although this finite-size behavior could be used as a criterion to distinguish 
both phases in principle, it is not reliable generally as
it does not hold when $V$ is kept fixed and $U$ varies through the critical value $U_c$.
Moreover, this is a transient behavior for intermediate values of $1/L$ only as
$D_{\rho}(L) \approx A \exp(- L/\xi)$ is also a convex function of $1/L$ for large 
enough $L \gg \xi$ in the insulating phase.

The Luttinger parameter $K_{\rho}$ also extrapolates to a finite value 
in the insulating phase if the available system sizes are too small.\cite{ejima} 
Nevertheless, the direct calculation of $K_{\rho}$ from the
finite-size scaling of correlation functions
allows one to determine the metal-insulator phase boundary 
very accurately
because this approach can be applied for open boundary conditions and
thus with DMRG one can simulate systems which are one to two orders of 
magnitude larger than with periodic boundary conditions.

\section{Summary}
\label{sec:summary}

We have studied the transport properties of the $t$-$U$-$V$ extended Hubbard
model 
at quarter-filling by using the DMRG technique combined with a 
variational principle for the Drude weight. 
The contour map of the charge and spin Drude weights
has been determined 
in the parameter space of the on-site Coulomb repulsion $U$ and
nearest-neighbor Coulomb repulsion $V$.
We have found that (i) both Drude weights decrease monotonically
with increasing Coulomb repulsion, (ii) the charge Drude weight is
larger than the spin Drude weight in the metallic phase (Luttinger liquid),
(iii) the charge Drude weight is discontinuous across the 
Kosterlitz-Thouless transition from the metallic phase to the CDW insulating
phase. 

We have also discussed the finite-size scaling and the extrapolation to 
the thermodynamic limit of our numerical data.
In the insulating phase we find a universal relation between the charge gap 
and the correlation length which controls the exponential decay of finite-size
corrections to the charge Drude weight. Unfortunately, we reach the conclusion that
it is difficult to determine the phase boundary of a metal-insulator 
transition using the Drude weights calculated with DMRG because this approach
requires periodic boundary
conditions which considerably reduce the performance of DMRG and thus the
available system sizes for the finite-size scaling analysis.

\begin{acknowledgments}
We thank S. Ejima and S. Nishimoto for useful discussions. 
This work was supported in part by Grants-in-Aid for 
Scientific Research from the Ministry of Education, 
Science, Sports and Culture of Japan (No. 19$\cdot$963).  
A part of the computations was carried out at the Research Center for 
Computational Science, Okazaki Research Facilities, and 
the Institute for Solid State Physics, University of Tokyo.  
\end{acknowledgments}


\begin{thebibliography}{99}

\bibitem{zotos} X. Zotos, J. Phys. Soc. Jpn. Suppl. {\bf 74}, 173 (2005). 
\bibitem{kubo} R. Kubo, J. Phys. Soc. Jpn. {\bf 12}, 570 (1957). 
\bibitem{kohn} W. Kohn, Phys. Rev. {\bf 133}, A171 (1964). 
\bibitem{jerome} D. J\'erome, Science {\bf 252}, 1509 (1991). 
\bibitem{seohotta} H. Seo, C. Hotta, and H. Fukuyama, Chem. Rev. {\bf 104}, 5005 (2004). 
\bibitem{lorenz} T. Lorenz, M. Hofmann, M. Gr\"uninger, A. Freimuth, G. S. Uhrig, M. Dumm, 
and M. Dressel, Nature {\bf 418}, 614 (2002). 
\bibitem{schulz} H.J. Schulz, {\it Strongly Correlated Electronic Materials}, 
ed. K. S. Bedell, Z. Wang, D. E. Meltzer, A. V. Balatsky and E. Abraham 
(Addison-Wesley Publish Company, 1994) p. 187. 
\bibitem{solyom} J. S\'olyom, Adv. Phys. {\bf 28}, 201 (1979). 
\bibitem{emery} V.J. Emery, in {\it Highly Conducting One-Dimensional Solids}, 
edited by J.T. Devreese, R. Evrand, and V. van Doren 
(Plenum, New York, 1979), p. 327.
\bibitem{voit} J. Voit, Rep. Prog. Phys. {\bf 58}, 977 (1995). 
\bibitem{giamarchibook} T. Giamarchi, {\it Quantum Physics in One dimension}, 
(Oxford Science Publications, 2004), p.391. 
\bibitem{sologubenko} A.V. Sologubenko, K. Gianno, H.R. Ott, U. Ammerahl, 
and A. Revcolevschi, Phys. Rev. Lett. {\bf 84}, 2714 (2000).
\bibitem{hess} C. Hess, C. Baumann, U. Ammerahl, B. B\"uchner, 
F. Heidrich-Meisner, W. Brenig, and A. Revcolevschi, 
Phys. Rev. B {\bf 64}, 184305 (2001). 
\bibitem{kudo} K. Kudo, S. Ishikawa, T. Noji, T. Adachi, Y. Koike, 
K. Maki, S. Tsuji, and K. Kumagai, 
J. Phys. Soc. Jpn. {\bf 70}, 437 (2001).
\bibitem{meisner} F. Heidrich-Meisner, A. Honecker, and W. Brenig, 
Eur. Phys. J. Special Topics {\bf 151}, 135 (2007). 
\bibitem{poilblanc} D. Poilblanc, T. Ziman, J. Bellissard, F. Mira, and G. Montambaux, 
Europhys. Lett. {\bf 22}, 537 (1993).
\bibitem{stafford} C. A. Stafford, A. J. Millis, and B. S. Shastry, 
Phys. Rev. B {\bf 43}, 13660 (1991); C. A. Stafford and A. J. Millis, 
Phys. Rev. B {\bf 48}, 1409 (1993).
\bibitem{bosoni} A. Luther and I Peschel, Phys. Rev. B {\bf 9}, 2911 (1974); 
D. C. Mattis, J. Math. Phys. {\bf 15}, 609 (1974); 
J.W. Cannon and E. Fradkin, Phys. Rev. B {\bf 41}, 9435 (1990); 
J. Voit, Phys. Rev. B {\bf 45}, 4027 (1992).
\bibitem{yoshioka1} H. Yoshioka, M. Tsuchiizu, and Y. Suzumura, 
J. Phys. Soc. Jpn. {\bf 69}, 651 (2000). 
\bibitem{yoshioka2} H. Yoshioka, M. Tsuchiizu, and Y. Suzumura, 
J. Phys. Soc. Jpn. {\bf 70}, 762 (2001). 
\bibitem{fourcade} B. Fourcade and G. Spronken, Phys. Rev. B {\bf 29}, 5089 (1984).
\bibitem{andergassen} S. Andergassen, T. Enss, V. Meden, W. Metzner, U. Schollw\"ock, 
and K. Sch\"onhammer, Phys. Rev. B {\bf 73}, 045125 (2006).
\bibitem{edcal} L. M. del Bosch and L. M. Falicov, Phys. Rev. B {\bf 37}, 6073 (1988); 
B. Fourcade and G. Spronken, Phys. Rev. B {\bf 29}, 5096 (1984).
\bibitem{milazotos} F. Mila and X. Zotos, Europhys. Lett. {\bf 24}, 133 (1993). 
\bibitem{penc} K. Penc and F. Mila, Phys. Rev. B {\bf 49}, 9670 (1994). 
\bibitem{nakamura} M. Nakamura, Phys. Rev. B {\bf 61}, 16377 (2000). 
\bibitem{sano} K. Sano and Y. \^Ono, Phys. Rev. B {\bf 70}, 155102 (2004). 
\bibitem{qmccal} J.E. Hirsch, Phys. Rev. Lett. {\bf 53}, 2327 (1984); 
J.E. Hirsch and D.J. Scalapino, Phys. Rev. B {\bf 27}, 
7169 (1983); {\bf 29}, 5554 (1984); 
H.Q. Lin and J.E. Hirsch, Phys. Rev. B {\bf 33}, 8155 (1986); 
J.W. Cannon, R.T. Scalettar, and E. Fradkin, 
Phys. Rev. B {\bf 44}, 5995 (1991).
\bibitem{schmitteckert} P. Schmitteckert, R. Werner, Phys. Rev. B {\bf 69}, 195115 (2004). 
\bibitem{ejima} S. Ejima, F. Gebhard, and S. Nishimoto, 
Europhys. Lett. {\bf 70}, 492 (2005).
\bibitem{lieb} E.H. Lieb and F. Y. Wu, Phys. Rev. Lett. {\bf 20} 1445 (1968). 
\bibitem{halden} F.D.M. Haldane, Phys. Rev. Lett. {\bf 45}, 1358 (1980); 
{\bf 47}, 1840 (1981); J. Phys. C {\bf 14}, 2585 (1981). 
\bibitem{yangyang} C.N. Yang and C. P. Yang, Phys. Rev. {\bf 150}, 321 (1966), 327.
\bibitem{fowler} M. Fowler and M. W. Puga, Phys. Rev. B {\bf 18}, 421 (1978). 
\bibitem{suzumura} Y. Suzumura, J. Phys. Soc. Jpn. {\bf 66}, 3244 (1997). 
\bibitem{chandra} P. Chandra, P. Coleman, and A. I. Larkin, J. Phys.: 
Condens. Matter {\bf 2}, 7933 (1990). 
\bibitem{kopietz} P. Kopietz, Phys. Rev. B {\bf 57}, 7829 (1998). 
\bibitem{shastry} B.S. Shastry, and B. Sutherland, 
Phys. Rev. Lett. {\bf 65}, 243 (1990). 
\bibitem{millis} A. J. Millis and S. N. Coppersmith, 
Phys. Rev. B {\bf 42}, 10807 (1990). 
\bibitem{fye} R. M. Fye, M. J. Martins, D. J. Scalapino, J. Wagner, and W. Hanke
Phys. Rev. B {\bf 44}, 6909 (1991). 
\bibitem{kopietzheisen} P. Kopietz, Mod. Phys. Lett. B {\bf 7}, 1747 (1993). 
\bibitem{eric2002} E. Jeckelmann, Phys. Rev. B {\bf 66}, 045114 (2002). 
\bibitem{book08} E. Jeckelmann and H. Benthien in \textit{Computational Many Particle Physics}, 
H. Fehske, R. Schneider, and A. Wei\ss e (Eds.), Lecture Notes in Physics 739, pp. 621-635, 
Springer-Verlag, Berlin, Heidelberg, 2008. 
\bibitem{eric2003} E. Jeckelmann, Phys. Rev. B {\bf 67}, 075106 (2003).
\bibitem{carvalho} F.C. Dias, I.R. Pimentel, and M. Henkel, 
Phys. Rev. B {\bf 73}, 075109 (2006). 
\bibitem{giamarchi1} T. Giamarchi, Phys. Rev. B {\bf 44}, 2905 (1991). 
\bibitem{giamarchi2} T. Giamarchi, Phys. Rev. B {\bf 46}, 342 (1992). 
\bibitem{chubukov} A.V. Chubukov, D.L. Maslov, and F.H.L. Essler, 
Phys. Rev. B {\bf 77}, 161102(R) (2008).
\bibitem{scalapino} D.J. Scalapino, S.R. White, and S. Zhang, 
Phys. Rev. B {\bf 47}, 7995 (1993). 
\bibitem{rigol} M. Rigol and B.S. Shastry, Phys. Rev. B {\bf 77}, 161101(R) (2008). 
\end{thebibliography}
\end{document}